\documentclass[10pt,conference]{IEEEtran}
\IEEEoverridecommandlockouts
\usepackage{cite}
\usepackage{amsmath,amssymb,amsfonts}
\usepackage{algorithmic}
\usepackage{graphicx}
\usepackage{textcomp}
\usepackage{xcolor}
\def\BibTeX{{\rm B\kern-.05em{\sc i\kern-.025em b}\kern-.08em
  T\kern-.1667em\lower.7ex\hbox{E}\kern-.125emX}}
\begin{document}

\title{A Reference Architecture for Blockchain-based Traceability Systems Using Domain-Driven Design and Microservices \\
}

%\author{\IEEEauthorblockN{Anonymous Authors}}

% \author{\IEEEauthorblockN{1\textsuperscript{st} Yanze Wang}
% \IEEEauthorblockA{\textit{Software Institute} \\
% \textit{Nanjing University}\\
% Nanjing, China \\
% yzwang9826@gmail.com}
% \and
% \IEEEauthorblockN{2\textsuperscript{nd} Shanshan Li}
% \IEEEauthorblockA{\textit{Software Institute} \\
% \textit{Nanjing University}\\
% Nanjing, China \\
% stu.shanshan.li@gmail.com}
% \and
% \IEEEauthorblockN{3\textsuperscript{rd} Huikun Liu}
% \IEEEauthorblockA{\textit{Software Institute} \\
% \textit{Nanjing University}\\
% Nanjing, China \\
% chloeliuliu@126.com}
% \and
% \IEEEauthorblockN{4\textsuperscript{th} He Zhang}
% \IEEEauthorblockA{\textit{Software Institute} \\
% \textit{Nanjing University}\\
% Nanjing, China \\
% hezhang@nju.edu.cn}
% \and
% \IEEEauthorblockA{\textit{???} \\
% \textit{Huawei Technologies Co., Ltd.}\\
% Nanjing, China \\
% panbo17@huawei.com}
% % \and
% % \IEEEauthorblockN{6\textsuperscript{th} Given Name Surname}
% % \IEEEauthorblockA{\textit{dept. name of organization (of Aff.)} \\
% % \textit{name of organization (of Aff.)}\\
% % City, Country \\
% % email address or ORCID}
% }

\author{\IEEEauthorblockN{Yanze Wang$^{1}$, Shanshan Li$^{1,*}$, Huikun Liu$^1$, He Zhang$^1$, Bo Pan$^2$}
\IEEEauthorblockA{
$^1$State Key Laboratory of Novel Software Technology, Software Institute, Nanjing University, Nanjing, China\\ 
$^2$Terminal Cloud Service Department, Huawei Technologies Co.,Ltd, Nanjing, China\\
$^*$\textit{Corresponding Author}\\
Email: \{yzwang9826@gmail.com, lss@nju.edu.cn, chloeliuliu@126.com, hezhang@nju.edu.cn, panbo17@huawei.com\}
}}

\maketitle

\begin{abstract}
Traceability systems are important for solving problems due to the increasing scale of the global supply chain, such as food safety crises and market disorder. Blockchain, as an immutable and decentralized ledger, is able to optimize the traditional traceability system by ensuring the transparency and reliability of the system data. However, the use of blockchain technology may lead to a rapid increase in the complexity of system design and development. It is challenging to address widespread and complicated business, changeable processes, and massive data in practice, which are the main factors restricting the wide application of a blockchain-based traceability system (BTS). Therefore, in this paper, we reviewed relevant studies and proposed a reference architecture for BTSs. The proposed reference architecture can improve the cohesiveness, maintainability, and extensibility of BTSs through domain-driven design (DDD) and microservices. Considering the efficiency reduction caused by massive data and complicated data structure, we further changed the traditional single blockchain framework into multiple sub-chain networks, which could improve development efficiency and system performance. With the guidance of the architecture trade-off analysis method (ATAM), we evaluated our reference architecture and implemented a prototype in the salmon supply chain scenario. The results show that our solution is effective and adaptable to meet the requirements of BTSs.
\end{abstract}


\section{Introduction}
\label{sec:intro}
%引出区块链为止尽量合并（内容缩写）；参考文献补充；第一句点明主旨
%Despite growing concerns about food safety and quality in the global supply chain industry, 
Numerous food safety scandals have severely undermined consumer confidence in the food industry and have caused growing concerns about food safety and quality in the global supply chain industry~\cite{rochefort2021rapid}. People are struggling to find an effective way to deal with these issues, and the term ``traceability system'' is now being used more frequently than ever before~\cite{olsen2018components}. The core concept of an ideal traceability system was described by Kim et al.\cite{kim1995ontology} in 1995, which is the ability to track products and activities. These activities include buying, selling, transporting, etc.~\cite{bevilacqua2009business}. All of these activities can describe the route of the product to solve the food safety crisis and increase people's confidence. As a result, the traceability system is gradually applied in all walks of life. However, there are still problems with traditional traceability systems that prevent their widespread use: (1) Data are often opaque and fabricated~\cite{zhang2020ColdChainLogistics}. (2) Most traceability systems are currently based on a centralized database, which is easy to manipulate~\cite{zhang2022research}. (3) Blocking data communications between enterprises and centralized databases leads to single points of failure and business bottlenecks\cite{Bodkhe2020review}. 

%关联 secIII A B
Blockchain~\cite{nakamoto2008bitcoin} is a popular solution for traceability systems, which combines cryptography, consensus mechanisms, and smart contracts to maintain data transparency and reliability. It can provide an immutable and transparent method for recording system transactions~\cite{queiroz2019blockchain}. In addition to the benefits, blockchain may also pose difficulties in designing, developing, and operating traceability systems. Currently, the architecture design of blockchain-based traceability systems (BTSs) is a hot but immature research topic~\cite{hastig2020blockchain}. First, the complexity and breadth of the current business process for BTSs~\cite{handayani2021blockchain} can lead to a sharp increase in the workload of the development, implementation, and maintenance of traceability systems, especially in smart contracts. Frequent changes in business affect the normal operation of the function and reduce the efficiency and precision of traceability, causing a demand for extensibility~\cite{zhao2020blockchain,adamashvili2021blockchain}. Second, the business process traceability system is complex, and the supply chain needs to involve many components like production, supply and marketing as well as the dependencies among them in each link, which increases system complexity greatly~\cite{Islam2021708}. Third, the large amount of business data leads to a degradation in the performance of blockchain storage, and the changeable data structure increases the overhead of blockchain data operation and maintenance~\cite{mirabelli2020blockchain}.
%\cite{zhao2020blockchain,adamashvili2021blockchain}
%\cite{qian2020food, mirabelli2020blockchain}. 

%we need to deal with two issues: first, to achieve a reasonable split of business scenarios, and second, to adopt a common approach to build the architecture for most business domains.

To address the challenges mentioned above, we conducted this in-depth research considering three motivations: (1) Explore a common approach to build the architecture for a complex BTS in the general business domain. (2) Create a reasonable division schema of business scenarios and delineate clear dependencies between domains to deal with possible business changes and domain complexity. (3) Study how to better maintain reliability under massive data. 

Specifically, \textit{for the first aspect}, we tend to propose a reference architecture that models the components, functions, and data flows of complex BTSs. The reference architecture can shape the overall quality of the system and can be further adjusted to business goals~\cite{kohler2019towards, cloutier2010concept}. Through the reference architecture, we can also ensure the standardization of the concrete system architectures, achieve system interoperability, decrease design complexity, and improve development productivity~\cite{haesevoets2014architecture,angelov2012framework}. Then, \textit{for the second and third aspects}, we utilize the domain-driven design (DDD) with microservices to realize the business scenario division and decrease domain complexity of a BTS. DDD is a software development method that can help developers design high-quality software models~\cite{evans2004domain} and is common to be roughly divided into four layers. It can establish and separate domain models through concepts such as domain objects and detailed hierarchical design~\cite{rademacher2020deriving}. DDD isolates the domain layer, and software developers can choose the better model and design schemes to maintain consistency between the code implementation and the business model~\cite{rademacher2018challenges}. Additionally, through domain division, transactions can be processed in parallel in different domains, which can improve overall performance and data workload. Microservice architecture (MSA) is an architectural solution that divides the traditional single architecture into multiple services based on business modules. MSA can significantly improve development, operation, extensibility, and maintenance efficiency~\cite{ray2020extending, zhang2019microservice}. Moreover, MSA has gradually become a common way to realize a bounded context in DDD. In contrast to DDD's focus on dealing with complex domain, MSA focuses on independent development, testing, building, and deployment of architecture modules.

% For all the above reasons, we propose a reference architecture of the traceability system based on domain-driven design, blockchain, and microservices. 

The contributions of this paper are as follows.
\begin{itemize}
    \item We provide a reference architecture suitable for the design of complex BTSs. The proposed architecture has better maintainability and extensibility and can be widely used in many traceability domains.
    \item We introduce DDD into the proposed reference architecture, which controls the complexity caused by complicated and variable business scenarios and the dependencies among different domains, and improves the development efficiency of the system.
    \item We provide a method to decompose BTSs according to the business domain and implement a multi-chain as microservices, which improves the data processing capacity compared to the single chain.
    \item We evaluate our reference architecture following the architecture trade-off analysis method (ATAM) and implement a prototype in the salmon supply chain scenario. It not only displays the effectiveness of our solution but also provides a guide for researchers of other related work in the future. 
\end{itemize}

The paper is organized as follows. Section~\ref{section: Background and Related Work} gives background and related work. Section~\ref{section: Generation of Architecture} presents the generation of our reference architecture. Section~\ref{section: Reference Architecture} describes in detail the proposed reference architecture. Section~\ref{section: evaluation} is about the evaluation of our solution using ATAM. Section~\ref{TTV} discusses threats to validity. 
Finally, Section~\ref{section: conclusion} concludes the paper.

\section{Background and Related Work}\label{section: Background and Related Work}

%尽量缩写

\subsection{Blockchain-based Traceability Systems}
% Blockchain is the distributed ledger technology first outlined in 2008 by Nakamoto~\cite{nakamoto2008bitcoin}. Participants on chain can transact with each other directly in a secure, immutable and chronological way without the need for any intermediaries~\cite{lakhani2017truth, tonnissen2020analysing, swan2015blockchain}. 
% %Transactions is recorded in blocks and sorted on the block chain by transaction time. 
% The chain is stored by every participant, and the subsequent block is appended to the end of the chain, while maintaining the hash of the previous block, which keeps the on-chain data decentralized, immutable, and transparent. Such features are important to the traceability system, and problems such as opaque and fabricated data, malicious falsification, and single point of failure can be solved well. \par
%针对区块链溯源系统架构有所研究。。。
Many studies have explored the architecture design of Blockchain-based Traceability Systems (BTSs). For example, Chen et al.~\cite{kambilo2022drug} presented their framework in the traceability of drugs and developed a blockchain-based prototype. Li et al.~\cite{Li2021multi-chain} proposed a master-slave multi-chain blockchain in seafood product traceability system architecture, which improves the isolation of private data and solves the problem of poor load on a single blockchain. Mueen et al.~\cite{uddin2021blockchain} summarized the problems in the drug supply chain and proposed two different drug traceability systems based on Hyperledger Fabric and Hyperledger Besu frameworks. Xu et al.~\cite{xu2019designing} applied blockchain in a real-world project: OriginChain, for cross-border traceability.
However, the current effort still suffers from complicated blockchain-based framework and cannot balance between the extensibility and efficiency. There are still shortages in the development and maintenance of BTSs.

\subsection{Reference Architecture}
%升华
%缩写 
A reference architecture is a reference model mapped to system components and data flows between them~\cite{angelov2012framework}, which can provide clear descriptions of the interrelationships of complex systems and ensure the consistency of domain-specific solutions~\cite{cloutier2010concept}. By reducing the complexity of the system architecture, reference architectures are very attractive in complicated systems, especially when the organizations of the system become large and distributed~\cite{muller2010researching}.\par

Many studies have been carried out in the area of reference architecture.
Ataei et al.\cite{ataei2021neomycelia} provided an event-driven microservice architecture and has a good degree of applicability. Geest et al.\cite{van2021design}, by applying a domain-driven architecture approach, designed a reference architecture that could track and trace goods in Industry 4.0. Bhattacharya et al.\cite{bhattacharya2022blockchain} proposed a blockchain-based reference architecture for the chemical industries, introducing traceability and transparency, as well as lowering costs in the chemical product supply chain. A reference architecture for chain-wide transparency systems in meat supply chains was presented in the study by Kassahun et al.\cite{kassahun2016realizing}. The work of Isaja et al.\cite{isaja2017combining} introduced a reference architecture for industrial automation that takes advantage of edge computing and blockchain technologies. \par

\subsection{Domain-Driven Design and Microservices}
%概述
DDD is a method for the domain analysis and modeling of complex software systems. The complexity of software engineering arises from complicated domain business and design~\cite{ashfaq2021intelligent}, which are difficult to address using traditional approaches. Evans et al.\cite{evans2004domain} proposed the four-layer architecture of DDD, including the presentation, application, domain and infrastructure layer, and is widely used in numerous scenarios. \par

%b.战略设计、战术设计
DDD includes two processes, strategic design and tactical design. \textit{Strategic design} is a method to decompose the problem space, focusing on model decomposition and complexity control~\cite{vernon2013implementing}. The bounded context, an important component of strategic design, is a division of the solution space. Components within a bounded context have common responsibilities, and the semantics within that scope are explicit. \textit{Tactical design} is the detailed design and implementation of the domain model in each bounded context. An entity is an object with an identity, and its business changes are tracked during the tactical design process. Domain events represent events that occur in the problem space that are of interest to domain experts, mainly for documenting model history and communication across boundaries. Repositories encapsulate the persistence operations of domain entities in the infrastructure layer, decoupling the domain layer from the infrastructure layer. \par

% And a context map is a drawing that documents bounded contexts and their relationships. The user stories are the specific functions that the user must perform at each stage.\par
%c.战术设计

%Microservice 
The microservice architecture is a popular architectural style that supports the extensibility and maintainability of systems~\cite{zhang2019microservice}. Each small and independent microservice contains its own business logic, user-handling functions and back-end functions, and the most attractive characteristic is the decomposition of complex applications for better development and maintenance. Microservices are autonomous and communicate by open protocols, hence, they can be developed independently and even with different technologies~\cite{li2021understanding}.

%优势总结
% By introducing DDD and microservices, we can alleviate the poor maintainability and extensibility problems of BTSs. In addition, 

This paper provides a reliable reference architecture by combining DDD and microservices to improve the BTS design. Each microservice in our architecture has better cohesion, maintainability, and extensibility and can cope with a complex business framework and variable requirements.

\section{Generation of Architecture}\label{section: Generation of Architecture}
\subsection{Problem Identification }
%提了一下rapid review的版本
% In order to have a clearer understanding of panorama and better identification of industry issues, we further conducted a rapid review, following the guidelines presented by Kitchenham et al.\cite{keele2007guidelines}\par
% Our aim is to identify common problems and requirements with the traceability system, so we formulated our research questions as: (1) The application scenario that study focuses on. (2) the business logic and domain requirements of the application scenario. (3) The architecture to realize the functional and non-functional requirements. And (4) the implementation and evaluation of the architecture \par
% Through rapid review, we collect the result and answer the research questions: We find that traceability systems based on blockchain are widely used in many application scenarios, such as the food industry, seafood logistics, manufacturing industry, agriculture etc. The business models and requirements in each domain are completely different, which leads to the different design solutions and implementations of the architectures. \par
% These results describe a fact that the common method to build an architecture for blockchain traceability system is still lack extensive studies, and the architecture designs of blockchain traceability system are still immature. As we have mentioned in introduction section, business complexity and massive data have become the two major constraints. We start by detailing these two problems to better understand the requirements on the artifact.\par

%直接说明问题的版本:literature research 
 
To get a clearer picture of BTS design, we conducted literature research and found that BTSs are widely used in many application scenarios~\cite{dutta2020blockchain, handayani2021blockchain}, such as the aquaculture~\cite{parreno2014advanced, Li2021multi-chain}, cold-chain industry~\cite{masudin2021traceability}, pharmaceutical industry~\cite{uddin2021blockchain, kambilo2022drug}, agriculture~\cite{yang2021trusted, saurabh2021blockchain}%\cite{yang2021trusted, tian2016agri, Chen2021agri, saurabh2021blockchain}
, etc. Each domain has different business models and requirements, leading to different architectural solutions and design implementations~\cite{mirabelli2020blockchain, qian2020food}.
% manufacturing industries~\cite{eryilmaz2020manufacturing}

The literature review reflects that the architecture designs of the BTS are immature~\cite{hastig2020blockchain} and it still lacks a common and powerful architecture design solution~\cite{varavallo2022traceability}. It is difficult to ensure the overall performance, extensibility, and maintainability of systems~\cite{zhao2020blockchain,adamashvili2021blockchain} with different scenarios. Each business manages its own system, leading to mismatches between software and data structures\cite{khan2020iot}. As mentioned in Section~\ref{sec:intro}, business complexity and massive data have become the two main constraints. We describe these two problems to better understand the requirements as follows.

The first problem is the complexity and breadth of the current business process for traceability systems.
In a traceability system scenario, products need to go through multiple stages. Different scenarios have different processes, and it is exhausting to customize a dedicated blockchain-based architecture for each scenario independently. Complicated processes also make traceability more challenging.
Additionally, there are many factors that affect business alignment. Frequent business variations lead to changes in the traceability process and data structure, resulting in difficulty in maintenance of smart contracts and traditional software components. Due to the one-time deployment and difficulty of modification features of smart contract, the normal operation of traceability functions is affected, and also reduces the efficiency and accuracy of data. 
Therefore, it is important to ensure that the architecture is extensible enough to deal with changeable scenarios, which also leads to a dramatic increase in the complexity of smart contracts running on the blockchain.\par

The second problem is the massive and volatile data. Business scenarios are diverse and extensive, with a large volume of data and a volatile data structure. The data saved on the blockchain includes not only business information on the enterprise side but also user data on the client side. The huge amount of data leads to degradation and overload of the performance of the blockchain network, which puts pressure on the operation and maintenance of decentralized storage.
Moreover, storing such data in a single blockchain may not be suitable due to the large storage space and frequent change of data structure. 
% Therefore, the traditional blockchain-based traceability solution cannot meet the demands of a rapidly growing business economy.\par

\subsection{Requirement Definition and Analysis}
Based on the above problems identified in current BTSs and the concept of DDD and microservices, we have defined three general requirements and corresponding solutions, which is the initial design of the reference architecture.\par

\textbf{Requirement 1.} \emph{The architecture method needs to be suitable enough to describe various scenarios.}\par

We need to identify common points to construct a common approach to describe various business scenarios. And based on the literature and our practical experience, we found three common business processes: (1) Logistics tracking: It monitors and records the route of products and provides complete tracking results. (2) Warehouse management: It is also another important part of warehouse operation, and a reasonable design can ensure reliable data. (3) Entity information management: It can keep records of supply chain participants.\par

\textbf{Requirement 2.} \emph{Architecture should deal with frequent changes in real business}\par

Based on the solution of requirement 1, we introduced the concept of DDD and microservice to deal with a changing business process. DDD divides the complex business into domains using strategic design and tactical design, and we can use different domains to construct various business processes. And microservice is a typical method to realize the bounded context of DDD. Considering the granularity of the service, the dependency relationship, and the division of the boundaries, various components of the system are realized for construction. \par

\textbf{Requirement 3.} \emph{Architecture should maintain reliability and performance under massive data}\par

When DDD divides the entire business into different domains, it also divides the data into different parts. Each domain needs to maintain its own data, which are relatively isolated in different domains, so data mixing can be prevented. And based on the theoretical concept of microservices, a multi-chain framework is also designed. In this way, load balancing is achieved between business domains, which improves the reliability of the traceability system under large workloads. The load-balancing mechanism can also increase overall performance as transactions in different domains are executed in parallel on different blockchains. \par

Based on the requirements with solution described above and the hierarchical architecture of DDD, we design our reference architecture, which is introduced in the next section.

\begin{figure*}[!htbp]
\centerline{\includegraphics[width=1\textwidth]{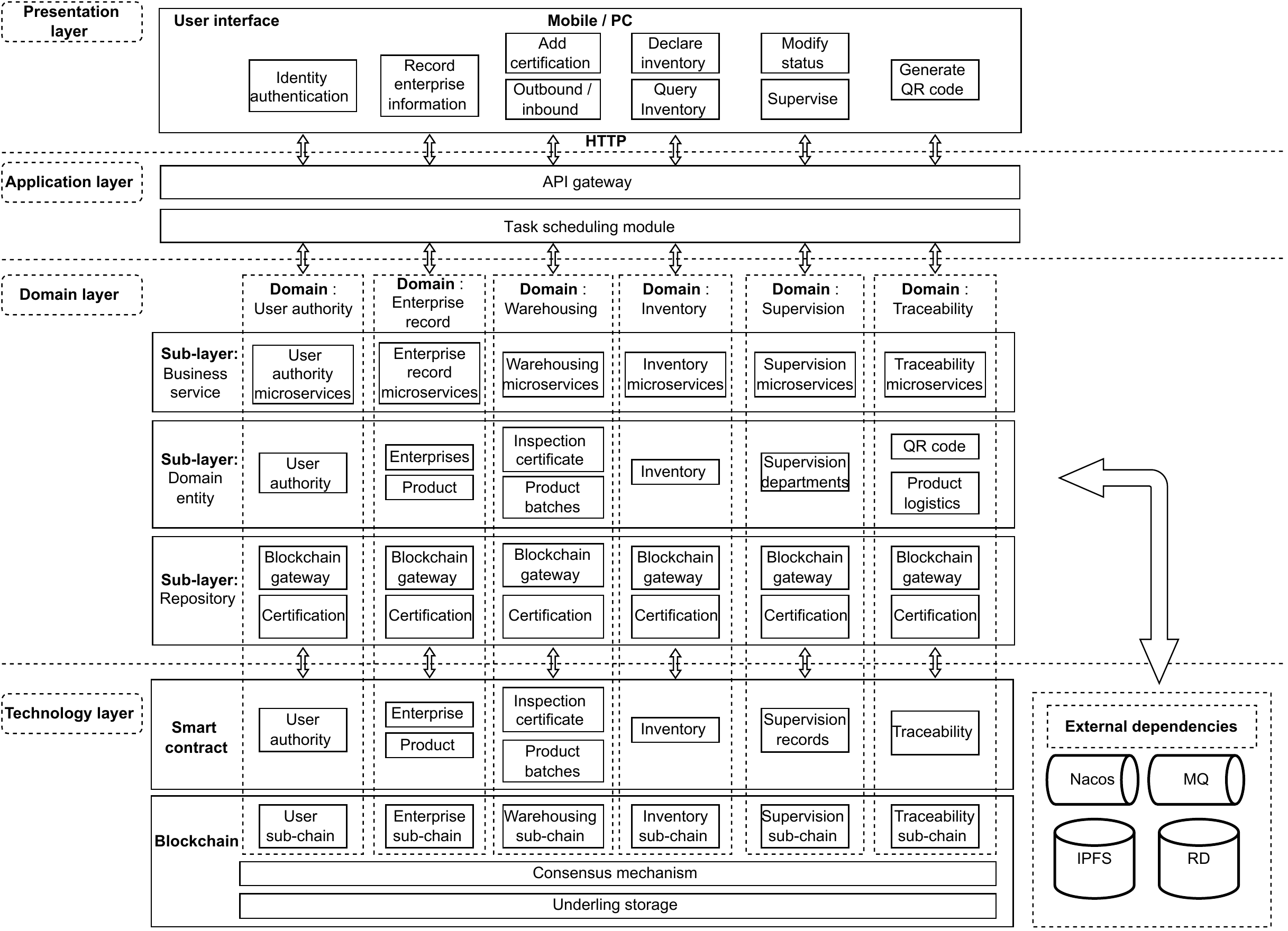}}
\caption{The Reference Architecture Proposed for BTSs}
\label{fig:Reference Architecture}
% \vspace*{-3ex}
\end{figure*}

\section{Reference Architecture for Traceability Blockchain system}\label{section: Reference Architecture}
 \subsection{Overall Framework}
%架构生成
%To better represent our design, we drew the overall system architecture, as shown in Fig~\ref{fig:Reference Architecture}. Our reference architecture describes the main components of the solution, the functionality they support, and their relationships, which includes three layers: application layer, domain layer and technology layer.
Based on the above section, we proposed a reference architecture for BTSs describing the main components, functions, and internal relationships, as shown in Fig.~\ref{fig:Reference Architecture}. DDD changes the traditional three-layer architecture into four-layer architecture, which are: presentation layer, application layer, domain layer, and technology layer. The presentation layer is used to handle requests and visual interface. The application layer is for domain model invocation. %The application layer is used to invoke the domain model and the repository for external services. The domain layer is the core of domain modeling and contains all of the business logic. The technology layer provides basic common services to the upper layers. 
In domain layer, we logically structure it into three sub-layers because they are key to expressing the core concept of business process and logic; other roles like value objects or domain events are included in these sub-layers.
At the bottom layer, we use the term ``technology'' instead of ``infrastructure'', because compared to the responsibility of data persistence, we introduced blockchain technology to ensure trusted, secure and decentralized data storage, combined with external dependencies, including middleware and database management systems. The functionalities are realized by smart contracts running on the blockchain, which can keep processes reliable while automating back-end processes.

%六个领域的生成
Besides the four layer structure , there are also include six domains in our reference architecture. According to our literature review, we have identified three common processes in Section~\ref{section: Generation of Architecture}, which are logistics tracking, storage management and entity information management. Then we split up the processes into single operations and got 25 different operations. Because each operation has its own concern, we can further cluster them into different domains based on the principle of separation concerns and context mapping patterns such as anti-corrosion layer, customer/supplier, and conformist proposed by Evans et al.\cite{evans2004domain} in DDD. And finally, we get six domains to describe the whole business as shown in Fig~\ref{fig:Reference Architecture}. 
%following the pattern proposed by Evans et al.\cite{evans2004domain}, we introduced context mapping patterns like anti-corrosion layer, customer/supplier pattern, and conformist pattern
%Following the separation concern principle and context mapping patterns such as anti-corrosion layer, customer/supplier, and conformist proposed by Evans et al.\cite{evans2004domain}, we started to identify the domain.
%We first separated the business framework from the application and technology and kept the framework aligned with the thinking of domain experts. We summarize three common processes in Section~\ref{section: Generation of Architecture}, which are logistics tracking, storage management, and entity information management. According to the principle of separation concerns in DDD, we divided them into six domains through separation and clustering of user stories based on the bounded context, in order to describe the business of the traceability system.

\textbf{Presentation layer:}
The presentation layer presents the necessary data information to users while receiving feedback from them, such as the graphical interface, user operation capture, data forwarding, etc., which is used to interact with an external user.
This layer consists of six functional parts that correspond to six domains. And the details of the domain and functionality division will be described below. All the necessary data in the below layers must be reflected in the presentation layer. The front end consists of mobile and PC terminals and is suitable for various situations.

\textbf{Application layer:}
The application layer is a thin layer that defines the tasks to be performed by the software and does not contain business rules or knowledge. It is responsible for organizing the flow of the entire application and coordinating the domain layer objects to perform the actual work. Here, we have designed it in two parts: API gateway and task scheduling module. The API gateway can communicate with presentation layer through HTTP, and task scheduling module's responsibility is to coordinate the work of the components such as domain entities, repository, etc. This layer usually accepts parameters from the presentation layer and then schedules domain entities and other components from the six domains to solve problems.

\textbf{Domain layer:}
The domain layer is the core layer of the DDD-based reference architecture.

As for the three sub-layers, the business service sub-layer is the business/domain logic of each domain, and it can publish or subscribe domain events and register service for the business demands. And the domain entity sub-layer consists of the participant entities including enterprises, common users, supervise departments, inventories, etc. Finally the repository sub-layer records the certification and relative information of the domain entities; it is also the way to communicate with the blockchain through its blockchain gateway.

And as for the six domains, the domain of user authority is mainly for user identity authentication, providing services to users, and blocking illegal access. Enterprises are the main participants in the supply chain, and the enterprise record domain is designed to keep detailed records for traceability and enterprise cooperation. The warehousing domain is to record the inbound and outbound products and then monitor and record the route of the product at all times. The inventory domain, such as transit stations or destination locations of the route, can identify the quantity and related information of the product in each inventory. The supervision domain keeps a close eye on the business entities and their operations and is the key to keeping the entire system controllable. The traceability domain is to record the route of products throughout the supply chain and provide timely feedback on the request for traceability information. Each domain is responsible for processing related business processes. The data accessed on-chain are constructed into domain entities, and the domain entities operate the business logic in them to provide services upward.

Among the six domains we mentioned above, there are three domains that are not the focus of our attention: the user authority domain, the enterprise record domain and the supervision domain. The domain of user authority and enterprise record are similar to the traditional access control system and member information management system. The supervision domain is mainly used for information modification and supervision, and its operations do not contain too much business logic. Therefore, we do not describe 
these three domain in detail. We discuss the four remaining types of domain and microservice in detail in the following section. 

\textbf{Technology layer:}
The main support technology that we use is blockchain, combined with other assistive technologies, including the InterPlanetary File System (IPFS), Nacos, etc. Here, we only introduce the general structure of a blockchain, and any blockchain that has a smart contract is suitable. Blockchain is used to store supply chain information about products and operations submitted by organizations in the supply chain. Smart contracts are deployed on each corresponding sub-chain, respectively, based on domain requirement; thus, there are six types of smart contract running on the blockchain.\par

Due to the large number of certification documents involved in each organization in the traceability system, we used IPFS as the system's file storage platform, which supports the on-chain and off-chain storage strategy to alleviate blockchain storage pressure. We introduce the QR code to record stocks and check products in the system, as it is ideal to accelerate inventory control~\cite{qian2020food}. Other information that does not affect business critical operations can be stored directly in the relational database (RD).\par

The system also introduces middleware that provides basic functions to ensure the integrity of the architecture, such as Nacos, a service registration and discovery middleware required for communication between microservices, enhancing the availability and reliability of the system. Moreover, the introduction of distributed message queues (MQs) provides the fundamental service for communication using domain events in each service, which can ensure timely publication and subscription of domain events in the system.\par

\subsection{Blockchain Network Architecture}
\begin{figure*}[!htbp]
\centerline{\includegraphics[width=1\textwidth]{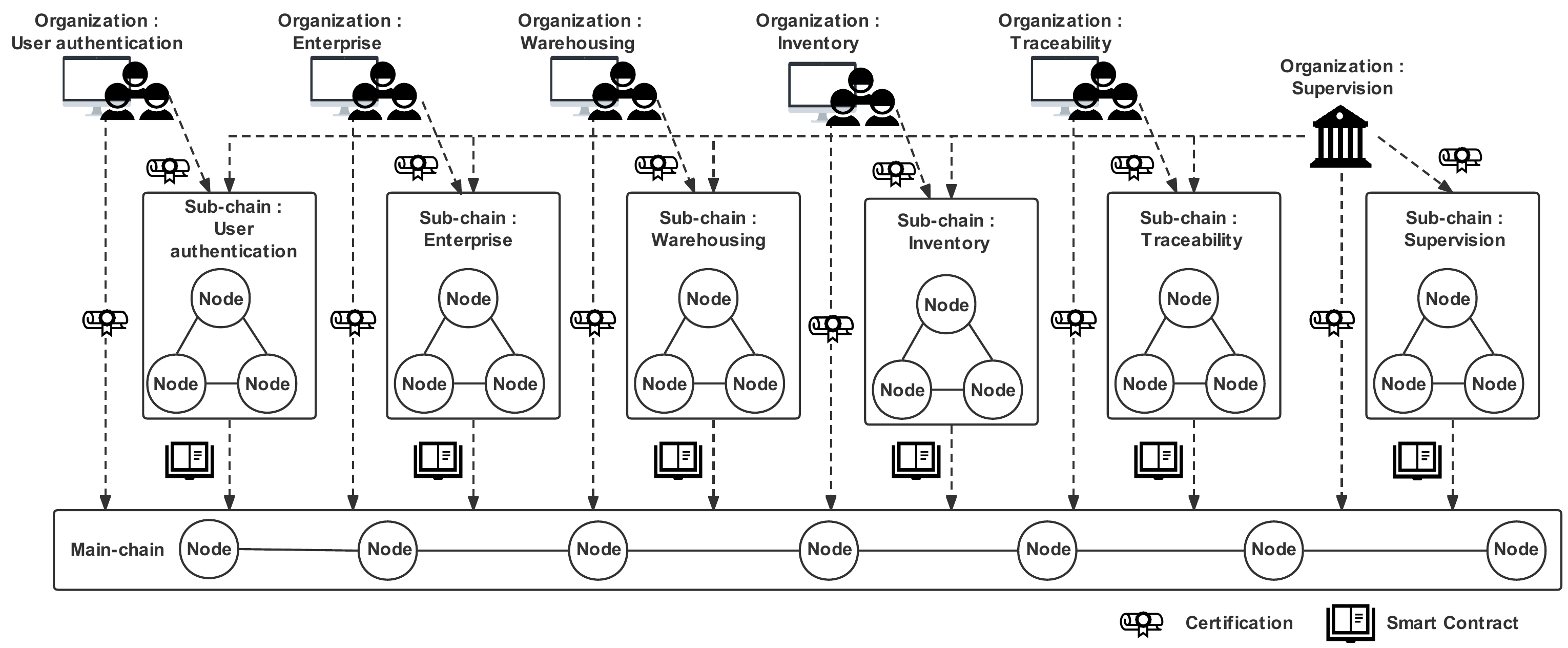}}
\caption{The Blockchain Network Architecture}
\label{fig:Blockchain Network}
% \vspace*{-3ex}
\end{figure*}

Microservice architecture is an architectural concept that divides complex systems into multiple modules. Each module focuses on one function, is highly cohesive internally but low-coupled. Based on the theoretical concept combined with blockchain technology, we establish a master-slave blockchain framework corresponding to the supply chain domains. Each sub-chain responsible for a specific domain including user authentication, enterprise, warehousing, inventory, traceability, and supervision of the supply chain. The reference architecture adopts a master-slave multi-chain storage model to manage traceability data and to construct a product traceability information management model, thus achieving overall monitoring of products in the supply chain. As shown in Fig.~\ref{fig:Blockchain Network}, based on the principle of microservice, the entire blockchain system is divided into six sub-chains with a main chain. Each sub-chain is registered with organizations that have access to the sub-chain's ledger by using the certification.These sub-chains undertake different business domain and connect to the main-chain through the smart contract.

% The warehousing sub-chain stores the inspection and quarantine certificate information with hash codes and batch information of product.
% The ledger of the inventory sub-chain stores the basic information of the inventories and the information of the products stored in the inventories.
% The traceability sub-chain records traceability information.
% The supervision sub-chain follows the supervisor's operation.
% The user authentication sub-chain stores the identities and certification of users.
% The enterprise sub-chain records relative information about the enterprises and the operations they performed.
The main chain only records metadata and indexes to facilitate data location when users and other entities send traceability requests. All companies in the supply chain are supposed to jointly maintain the main chain.

In this network, data privacy is protected. Only the organizations registered to the sub-chain have access to data in the ledger. Since the supervising organization has the demand of supervision, it is registered to all sub-chains. Only the public information of the product can be accessed by a traceability query and private data must be authorized by its owner.

Considering the large number of system users and transaction traffic, we are supposed to use non-Byzantine fault-tolerant consensus protocols such as Raft to improve performance. Besides, the world state database of blockchains should be as lightweight as possible, such as CouchDB. The smart contract design process should also try to avoid serious CPU-consuming operations and IO-intensive operations.

Smart contracts should be consistent with the granularity of domain entities for extensibility and modifiability. When the requirements of the system change, the developer only needs to modify the smart contract corresponding to the specific domain entity, without unnecessary upgrading of other contracts.

\subsection{Domain-Driven Design}
\emph{(1) Strategic Design}\par

Here, we will discuss the formation and mapping of bounded contexts. 
Based on the common process we have found, the user stories are identified and decomposed, and we initially obtain four user stories. Then we conduct further decomposition, until the sub-user stories are focused on only a single domain problem or become an operation. Next, each sub-user story is clustered according to the problem its concerns, and we obtain the six domains mentioned above. The bounded context is designed to define domain boundaries and we designed six bounded contexts, respectively.\par

%Besides, combined with business logic, following the pattern proposed by Evans et al.\cite{evans2004domain}, we introduced context mapping patterns like anti-corrosion layer, customer/supplier pattern, and conformist pattern in order to confirm the dependencies and relationship among these contexts.
In addition, combined with business logic, we introduced context mapping patterns to confirm the dependencies and relationships among these contexts.
We utilized the anticorrosion layer in the supervision context because of its dependence on multiple contexts including the enterprise context, the inventory context, and the warehouse context.
The warehousing context's operation will be recorded in the traceability context, which is a customer/supplier relationship.
The enterprise context relies on detailed information in the user context and the inventory context, so it forms a conformist relationship.\par

\emph{(2) Tactical Design}\par
Based on the above strategic design, we will discuss four domains in detail. For each domain, we outline its basic functionality and describe it through flowcharts.\par

\begin{figure}[htbp]
\centering
\includegraphics[width=0.9\linewidth]{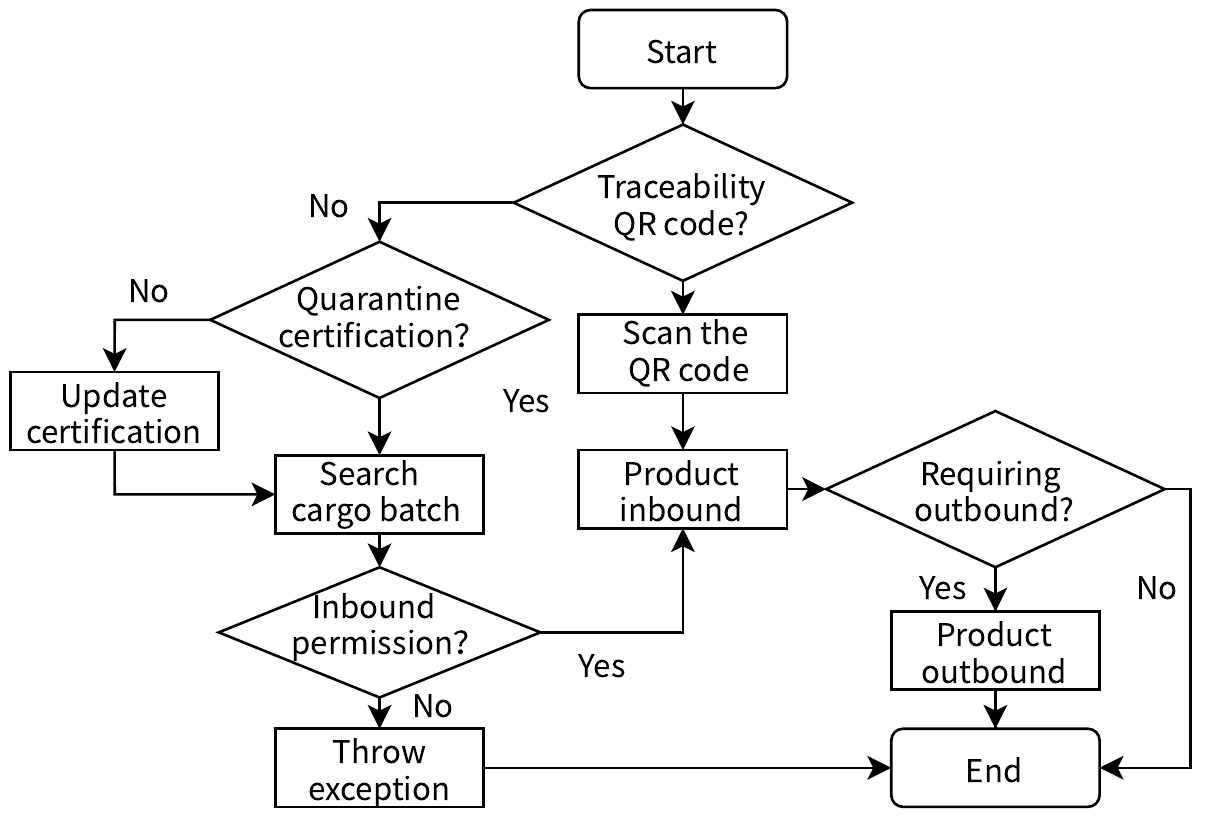}
\caption{The Key Process for the Warehousing Domain}
\label{fig:Warehousing Domain}
% \vspace*{-1ex}
\end{figure}

%In this section, we will discuss four domains in detail based on the result of domain division described above, which are the warehousing domain, the inventory domain, the enterprise record domain and the traceability domain. For each domain, we outline its basic functionality and describe it in detail through flowcharts.\par
%出入库领域
\textbf{The warehousing domain} is responsible for recording the inbound and outbound products.
The key process is shown in Fig.~\ref{fig:Warehousing Domain}. When the product is received from the upstream supplier, the enterprise first checks the receipt. If there is a QR code for traceability, the system will automatically inbound the product by scanning the QR code. Users of the enterprise need to search for the ID of the cargo batch through inspection and quarantine certifications because a certification ID is assigned to multiple batches of cargo that have been endorsed by the quarantine bureau. If there is no certification information for a certain batch of cargo, users should first update the relevant information in the system. \par
When users perform an inbound product operation, the warehouse will be checked according to the outbound information from the upstream supplier. If something is incorrect, an exception will be thrown. Otherwise, the warehousing domain will send warehousing events through the domain entity, then traceability domains and inventory domains will handle the warehousing events accordingly after monitoring them. After the outbound operation, the warehousing domain also sends domain events to the message queue.\par

\begin{figure}[htbp]
\centering
\includegraphics[width=0.8\linewidth]{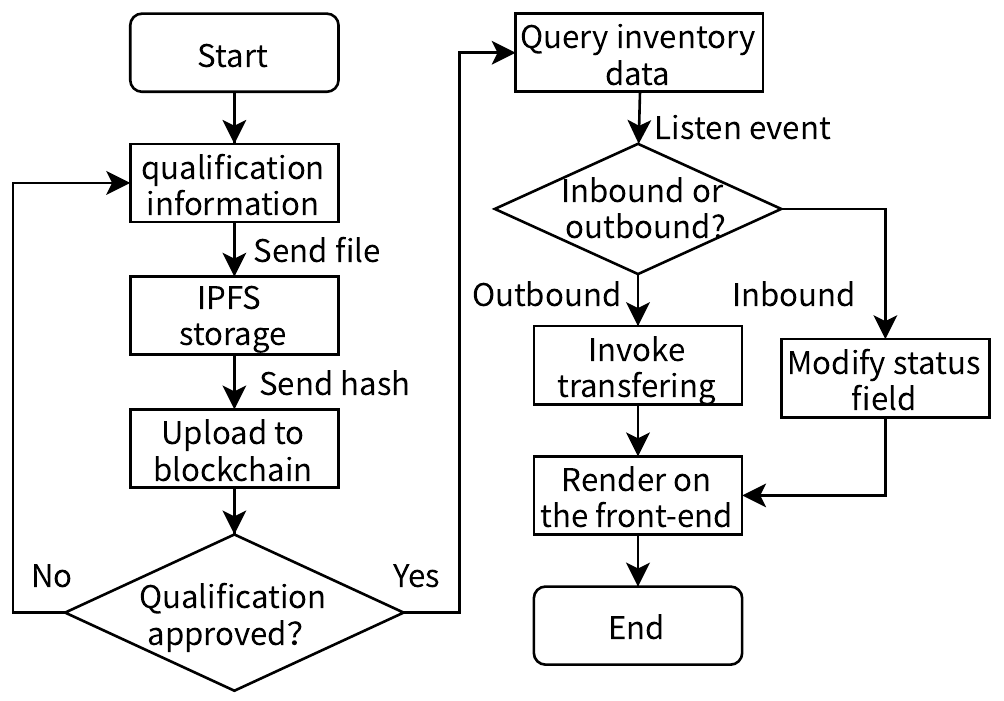}
\caption{The Key Process for the Inventory Domain}
\label{fig:Inventory Domain}
\vspace*{-1ex}
\end{figure}

%库存领域
\textbf{The inventory domain} serves the inventories and is responsible for the quantity of the product and related information. 
The inventory domain management flow chart is shown in Fig.~\ref{fig:Inventory Domain}. Here is the detailed process: the front-end receives the qualification uploaded by the user, which will be sent and stored in IPFS storage system of the back-end. After the file is successfully stored, an IPFS hash code of a specific proof file will be passed back to the front-end. When users submit the warehousing qualification information in the front-end, the hash code will be submitted together. The warehousing qualification information will be uploaded to the blockchain. When the warehousing qualification is approved by the supervision department, enterprise users can use the warehousing management function to query inventory data.

%The enterprise often operate the warehouse according to business demands, and the inbound and outbound service will send the inbound and outbound domain events timely. To be specific, 
The warehouse operation can trigger the sending of domain events, and the event listener in the inventory service listens to the inbound and outbound domain events in a timely way. If there is an outbound event, the inventory microservice invokes the smart contract on the blockchain inventory sub-chain to transfer the goods and assets. While it is an inbound event, the inventory microservice invokes the smart contract to modify the corresponding status field. All of the operations will be reflected on the front-end.

\begin{figure}[htbp]
\centering
\includegraphics[width=0.8\linewidth]{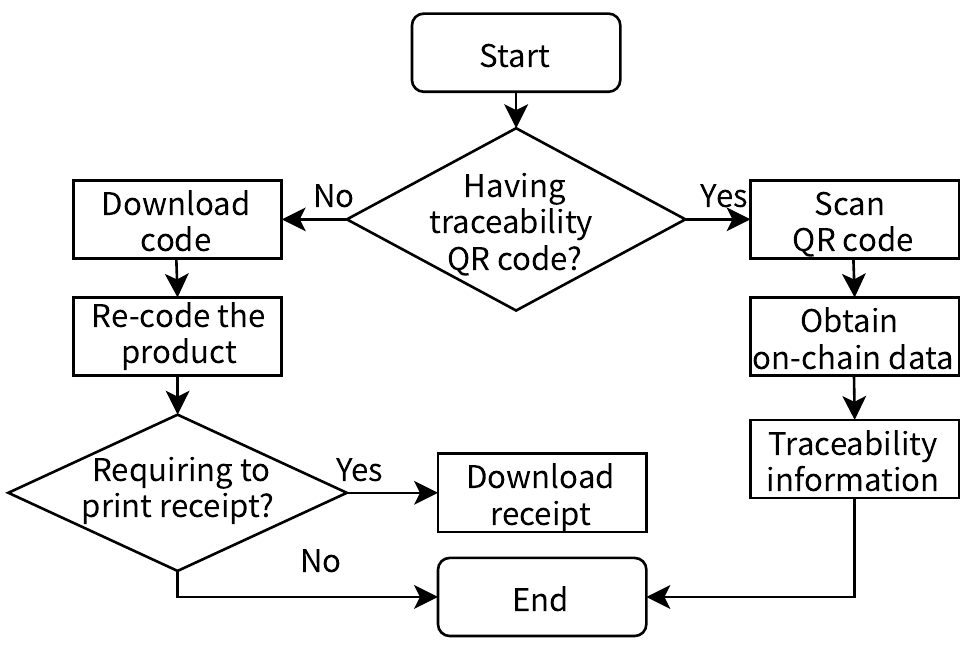}
\caption{The Key Process for the Traceability Domain}
\label{fig: Traceability Domain}
\vspace*{-1ex}
\end{figure}

\textbf{The traceability domain} access information on the route of products throughout the supply chain. The key process of the traceability domain is shown in Fig.~\ref{fig: Traceability Domain}. When organizations in the traceability supply chain receive a product, they first verify whether the package has a traceability QR code. If there is no QR code or it is damaged, the enterprise needs to download the QR code from the system and re-code the product. Enterprises can also print an outbound receipt with the QR code, which will be sent to the downstream company, where they can easily record the inbound product by scanning the QR code on the outbound receipt. Finally, any unit that gets a product with the correct QR code can be traced. When scanning the QR code in the package, the system immediately receives the traceability data on the blockchain and displays the result on the front end.\par

\par
In general, by combining the reference architecture with DDD and microservices, the system achieves better cohesiveness, maintainability, and extensibility. 
Each domain service can be run separately and deployed independently. When business changes, such as adding new logistics lines or introducing new companies, the functions of the system can be quickly expanded by releasing and integrating new services.
Since domain services are relatively isolated, transactions will be distributed and paralleled. For example, the traceability service focuses on querying the transaction process, while the warehousing service focuses on writing product and inventory information, which can improve system efficiency and availability.
In addition, due to the high cohesion of domain services, each team can focus on and choose the domain services they are good at for development and maintenance, improving agility and development efficiency. 

\section{Evaluation}\label{section: evaluation}
%中诚公司的介绍
%找ATAM的论文
Given the high level of abstraction in the reference architecture, it is a challenge to evaluate an abstract architecture~\cite{angelov2008contracting}. At this stage, we confirm and verify the viability and applicability of the reference architecture proposed by ATAM (architecture tradeoff analysis method)\cite{kazman1998architecture}. ATAM has been used for over a decade to evaluate architectures and is also a comprehensive approach to measure architecture's fitness with respect to multiple competing quality attributes.\par
Thus, we implemented a reference architecture prototype to introduce the ATAM. In general, the evaluation follows two phases. In phase 0, we need to clarify the partnership and prepare for informal meetings between stakeholders and architects and to work out the details of the architecture. In phase 1, everyone needs to begin to analyze what the system is about, including the overall architectural approaches, the quality attributes, the most typical scenes, and finally generate the document. Considering the necessity and space limitations, we are not going to talk about each detail of each step in ATAM, and all the key steps and results are shown below.\par

\subsection{Phase 0}
The evaluation was conducted with a blockchain research institute in China and applied to existing and new workflows. This research institute conducts research, development, and application of blockchain technology. More than 5,000 enterprises have been granted credit interaction on the platform, with more than 40 core independent intellectual property rights. The institute's advisory team includes experts and senior professors from home and abroad. And the core technical team of the institute includes technical experts from IBM, experienced system architects in blockchain technology. \par
We introduced them to ATAM and the stakeholders presented their business case. Due to China's import policy, companies are required to verify the traceability information of imported Norwegian salmon, and transactions need a series of traceability information. Therefore, we could demonstrate and evaluate the core functions through the case of ``The traceability information of imported Norwegian salmon''.\par

\subsection{Phase 1}
In this phase, we implemented a prototype of our reference architecture. Our prototype is based on the popular consortium blockchain platform, Hyperledger Fabric. Under the guidance of DDD and microservices, we analyzed the requirements of the seafood traceability system and obtained its core function, which was then combined with the quality requirement. Six bounded contexts of the system domain were identified and we further developed microservices, respectively. Spring Gateway was used to integrate various microservices. Additionally, we used IPFS to store business data to ensure security and performance. Finally, we also implemented the user interface for each role of the traceability system, which could ensure that the entire system was truly usable.\par
We then describe our architecture and quality attribute design to stakeholders and project managers to identify architecture approaches, build quality attribute utility trees, and analyze architectural approaches in priority scenarios.\par

\emph{1) Identifying architectural approach}

We use the IEEE standard for a software quality metrics methodology\cite{IEEE1061-1992, IEEE1061-1998} as a reference quality attributes model in our architecture design. According to the standard, %security is the ability that software can detect and prevent information leakage, information loss, illegal use and resource damage, while 
the extensibility is the effort required to improve or modify the efficiency and functionality of the software. To determine the architecture approach, we obtained information on performance and extensibility from the archetype.
\begin{itemize}
%\item For security, we discussed confidentiality, data leakage prevention, and permission control.
\item For performance, we realized that domain division and transactions can be processed in parallel in different domains.
\item For extensibility, we discussed the need for the system to have high extensibility to cope with business changes.
\end{itemize}
We explore the trade-offs, sensitivities, and risks of each quality attribute.

\emph{2) Utility tree elicitation}

Scenarios specify the stimuli that need to be captured and to which the architecture must respond. These captured stimuli are used to test the ability of the system to satisfy functional and non-functional requirements. In this step, we asked stakeholders to propose several different scenarios. Then we asked each stakeholder to vote on each scenario. In total, more than 25 scenarios were created, of which four have been selected. We describe those four scenarios as two journeys:
\begin{itemize}
\item Before shipping seafood, business users must record information about the current product with the upstream and downstream suppliers of the company in the system so that supervisors can track and review the products.
\item Business users, supervisors, and consumers can use the terminal device to obtain the traceability information of seafood by scanning the code on the product.
\end{itemize}\par

\begin{figure}[!htbp]
\centering
\includegraphics[width=0.9\linewidth]{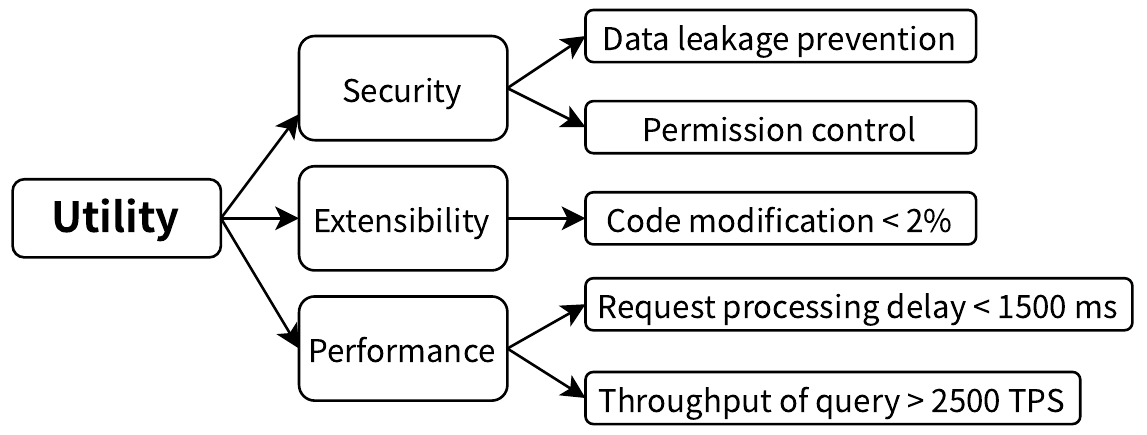}
\caption{The Utility Tree}
\label{fig:tree}
% \vspace*{-1ex}
\end{figure}

%改：补充三个标准的确定过程，以及描述
We have learnt that security, extensibility and performance are the most concerned attributes according to the stakeholders. 
For security, a portion of the data on-chain has a relatively high commercial value, and data leakage is unaffordable. Besides, different entities have different responsibilities , and illegal data access is forbidden no matter it is unintentional or malicious. Thus, security is crucial because it  affects the whole system.
For extensibility, there will definitely be business expansion or update in the future, so it is necessary to reduce the effort of code modification.
And for performance, lower latency and higher throughput can bring a better experience for customers, and can also facilitate data cooperation between enterprises.
Based on their feedback, we adjusted the quality attributes and generated a utility tree, as shown in Fig.~\ref{fig:tree}.
%We learned from stakeholders that security, extensibility, and performance were the quality attributes they were most concerned about. Therefore, we adjusted the quality attributes and generated a utility tree, as shown in Fig.~\ref{fig:tree}. 

\emph{3) Analyze architectural approaches} 

Based on the scenarios identified in the utility tree, we map our architectural approaches to business scenarios and quality attribute requirements and identify trade-off points.\par
Based on the salmon supply chain scenario, we imported real data into the prototype. Relevant components and events have also been created in the corresponding microservices of different domains. We also create different accounts for different roles in the traceability system, which can simulate the functions of a traceability system throughout the salmon supply chain to satisfy the demands of different typical scenarios. The three key quality attributes are discussed below.\par
\textbf{Security: }Our architecture is based on blockchain and IPFS storage, which are distributed and decentralized databases. The malicious have to invade more than half of the nodes in the system network to modify the record. In addition, all data are stored off-chain: only metadata, data hash, and references to data files are stored online; the malicious cannot find the data online directly. Also, different domains of data are stored in different databases and participants in different domains cannot access the data of others. The privacy and confidentiality of the data are well protected.\par
\textbf{Extensibility: }Through the combination of DDD and microservices, we divide the traceability business into several domains, and each domain in our architecture has better extensibility. We also design domain events with monitoring and handling functions to realize the decoupling of business logic and to facilitate cooperation between domain entities.\par
\textbf{Performance: }To have a better understanding of the performance, we conducted the test using the Hyperledger Caliper tool, which is a framework to benchmark the performance of blockchains for blockchain networks. The result shows that the querying throughput is about 2800 TPS, and the average latency is about 1.4s, which can satisfy the requirement. We use multiple sub-chains to deal with transactions parallel, which can balance load and reduces data storage and computing stress of each sub-chain.\par

There are two trade-off points in this reference architecture, i.e., the database system and the storage method. First, we use IPFS to store business data, which has better data security and privacy, but can significantly increase storage overhead compared to a traditional system like MangoDB. Second, our storage method is hash-storaged on-chain, whereas data are stored off-chain. This storage method keeps data confidential and complete, while there is a sacrifice in interoperability compared to the choice of storing data directly on the chain.\par

% \emph{4) Limitations}

% Although we realize better cohesion, maintainability and extensibility, it is still a fact that the architecture is more complex than these traditional software frameworks, and results in higher computing and storage overhead because of the use of multi-blockchains and IPFS storage mechanism. And our partnerships also agree with that it requires a lot of expertise. This reference architecture is more suitable for large-scale supply chain traceability scenarios with numerous roles and stakeholders.
\section{Threats to Validity}\label{TTV}
In the design and execution of this study, an extremely cautious attitude was taken to mitigate threats. Specifically, the main threats to validity and the corresponding mitigation strategies were considered when designing the reference architecture and conducting the evaluation.\par
\textbf{Conclusion Validity:} The threat to the validity of the conclusion could be the lack of expert evaluation. To mitigate this threat, we conducted the evaluation in authoritative blockchain research institude with a team of top experts, since interpreting the evaluation results and pointing out critical hidden facts often require experts' deep knowledge about the context.\par
\textbf{Internal Validity:} The architecture can behave differently in different scenarios and is more complex than traditional ones. To migrate this threat, we conducted sufficient research on primary studies and literature reviews and carefully designed our reference architecture to keep it reasonable.\par
\textbf{Construct Validity:} Experimenter bias occurs when researchers evaluate the architecture in their own perception. To migrate this threat, we reached a consensus with stakeholders to finalize quality attributes and generate a utility tree for objective evaluation.\par

\section{Conclusion and Future Work}\label{section: conclusion}
The development of global supply chains has increased the requirement for traceability, and the construction of a blockchain-based traceability system is a promising solution. However, this increases the difficulty and complexity of design, development, and operation. Therefore, we designed a reference architecture for a traceability system based on DDD, blockchain, and microservices, which control the complexity of system design, development, and operation. By utilizing multiple sub-chains, we improve the data processing capacity compared to a single-chain solution. Through a salmon BTS prototype and evaluation with stakeholders and business managers, we have demonstrated that our reference architecture meets quality requirements and can be widely used in many different scenarios. In the future, we plan to improve other QAs such as interoperability of our reference architecture, scale up the prototype, and achieve scenario customization for BTSs of different domains. Additionally, we will continue to perform a more solid evaluation with various real-world systems and iteratively improve our reference architecture.\par

\section*{Acknowledgment}

% The preferred spelling of the word ``acknowledgment'' in America is without 
% an ``e'' after the ``g''. Avoid the stilted expression ``one of us (R. B. 
% G.) thanks $\ldots$''. Instead, try ``R. B. G. thanks$\ldots$''. Put sponsor 
% acknowledgments in the unnumbered footnote on the first page.
This work is jointly supported by the National Key Research and Development Program of China (No.2019YFE0105500) and the Research Council of Norway (No.309494), the Key Research and Development Program of Jiangsu Province (No.BE2021002-2), the National Natural Science Foundation of China (No.62072227 and No.62202219), as well as the Intergovernmental Bilateral Innovation Project of Jiangsu Province (No.BZ2020017).

\bibliographystyle{IEEEtran}
\bibliography{IEEEabrv,ref}

\end{document}